\documentclass[aps,preprintnumbers,amsmath,amssymb,prb,superscriptaddress,longbibliography,twocolumn]{revtex4-2}
\usepackage{dcolumn}
\usepackage{color}
\usepackage[caption=false]{subfig}
\usepackage{feynmp}
\usepackage{feynmp-auto}
\usepackage{float}
\usepackage{bbm}
\usepackage{empheq}
\usepackage[dvipsnames]{xcolor}

\newcounter{mysubequations}

\renewcommand{\themysubequations}{(\roman{mysubequations})}

\newcommand{\mysubnumber}{\refstepcounter{mysubequations}\themysubequations}

\newcommand{\mycomment}[1]{}

\def\comment#1{}

\newcommand{\nc}{\newcommand}

\nc{\scs}{\scriptstyle}
\nc{\setval}{\fmfset{wiggly_len}{3mm} \fmfset{arrow_len}{1.5mm}
	\fmfset{arrow_ang}{13} \fmfset{dash_len}{1.5mm}\fmfpen{0.125mm}
	\fmfset{dot_size}{2thick}}

\usepackage{bm,latexsym,mathrsfs,enumerate,color}
\usepackage[mathcal]{euscript}
\usepackage[breaklinks=true,unicode=true,urlcolor = blue,colorlinks = true,citecolor = blue,linkcolor = blue]{hyperref}
\usepackage{graphicx}
\usepackage{todonotes}
\usepackage{wrapfig}

\renewcommand{\vec}[1]{\bm{#1}}

\def\slashchar#1{\setbox0=\hbox{$#1$}           
	\dimen0=\wd0                                 
	\setbox1=\hbox{/} \dimen1=\wd1               
	\ifdim\dimen0>\dimen1                        
	\rlap{\hbox to \dimen0{\hfil/\hfil}}      
	#1                                        
	\else                                        
	\rlap{\hbox to \dimen1{\hfil$#1$\hfil}}   
	/                                         
	\fi}                                         %

\DeclareMathAlphabet\mathbfcal{OMS}{cmsy}{b}{n}

\def\nablab{{\mbox{\boldmath $\nabla$}}}

\def\Pib{{\mbox{\boldmath $\Pi$}}}

\usepackage{physics}
\usepackage{xcolor}

\begin{document}
\title{The Feynman paradox in a spherical axion insulator}

\author{Anastasiia Chyzhykova}
\affiliation{Faculty of Physics of Dresden University of Technology, 01062 Dresden, Germany}
\affiliation{Institute for Theoretical Solid State Physics, IFW Dresden, Helmholtzstr. 20, 01069 Dresden, Germany}

\author{Jeroen van den Brink}
\affiliation{Institute for Theoretical Solid State Physics, IFW Dresden, Helmholtzstr. 20, 01069 Dresden, Germany}
\affiliation{Institute for Theoretical Physics and W\"urzburg-Dresden Cluster of Excellence ct.qmat, TU Dresden, 01069 Dresden, Germany}

\author{Flavio S. Nogueira}
\affiliation{Institute for Theoretical Solid State Physics, IFW Dresden, Helmholtzstr. 20, 01069 Dresden, Germany}

\begin{abstract}
We show that a small charged probe near a spherical topological insulator causes the latter to rotate around a symmetry axis defined by 
the center of the sphere and the position of the charge outside the latter. The rotation occurs when the distance from the charge to the center of the sphere is changed. This phenomenon occurs due to induced static fields and is a consequence of the axion electrodynamics underlying the electromagnetic response of a topological insulator. Assuming a regime where the charged probe can be regarded 
as a point charge $q=Ne$, where $N$ is a positive integer and $e$ is the elementary electric charge, we obtain that the rotation frequency 
is given by $\omega=(N\alpha)^2\Upsilon(\epsilon,d/a)/I$, where $I$ is the moment of inertia, $\alpha$ is the fine-structure constant, and the function $\Upsilon$ depends on the dielectric constant $\epsilon$ and the relative distance $d/a$ of the charge from the center of the sphere of radius $a$. Since the point charge also induces Hall currents on the surface, we also compute their associated angular momentum. This allows us to derive an exact expression for the electronic velocity on the surface as a function of $a/d$. 
\end{abstract}

\maketitle

\section{Introduction}

During the first years of activity in the field of topological insulators (TIs), 
it was predicted \cite{Qi-2008,Essin_2009,Ryu-axion} 
that three-dimensional TIs feature an electromagnetic response described by axion electrodynamics \cite{Wilczek}, whose 
Lagrangian in the absence of sources is given by
\begin{equation}
\label{Eq:Axion-ED}
\mathcal{L}=\frac{1}{8\pi}\left(\epsilon\vec{E}^2-\frac{1}{\mu}\vec{B}^2\right)-\frac{\alpha\theta}{4\pi^2}\vec{E}\cdot\vec{B},
\end{equation}
where $\alpha=e^2/(\hbar c)$ is the fine structure constant and $\theta$ is the axion, which is generally 
a field, but here is assumed uniform inside the insulator, vanishing otherwise \cite{Qi-2008,Ryu-axion}. The parameter $\theta$ is given by 
the flux of the Berry curvature associated with the band structure in momentum space \cite{Qi-2008}. For 
topologically nontrivial systems preserving either time-reversal (TR) 
or inversion symmetry, $\theta=\pi$ \cite{Ryu-axion}. Within the effective theory (\ref{Eq:Axion-ED}), a topologically trivial insulator has $\theta=0$, 
thus corresponding to a classical dielectric.

There is a large number of works dedicated to explore the consequences of the electromagnetic response described by Eq.~(\ref{Eq:Axion-ED}). 
One salient prediction refers to changes in the Faraday and Kerr rotation experiments \cite{Qi-2008,Karch}. This is an important 
experimental probe of the so called topological magnetoelectric effect, because results of order $\mathcal{O}(\alpha)$ can be obtained, which is 
under experimental reach \cite{Karch}. However, having $\theta=\pi$ is an experimental challenge, as it requires samples where the Fermi 
energy nearly vanishes, being precisely at the Dirac point. The topological magneotoelectric effect has been finally probed in Refs. 
\cite{Axion-exp-Armitage} (Bi$_2$Se$_3$) and \cite{Axion-exp-Molenkamp} (strained HgTe), thus confirming the theoretical prediction.  

The solution of the problem for a point charge $q$ in 
the presence of a topological (i.e., $\theta\neq 0$) dielectric sphere was previously discussed in the Supplemental Material of 
Ref. \cite{Qi_image_monopole} and in Ref. \cite{Martin-Ruiz_PhysRevD.94.085019}. Here we show that an important physical effect arises in this situation that was not considered so far: we demonstrate that the presence of a point charge imparts an electromagnetic angular momentum on the sphere, ultimately causing it to rotate as the position of the point charge is varied. Physically, the origin of the effect is simple: due to the axion term, the electric field generated by the point charge in the presence of the topological dielectric sphere induces a magnetic field and therefore the momentum of the electromagnetic field is nonzero \cite{Nogueira-van_den_Brink-Nussinov_London_axion,Nogueira-van_den_Brink_PhysRevResearch.4.013074}. As a consequence, an angular momentum is also induced by the electromagnetic field. In view of the conservation of the total angular momentum, this in turn generates a mechanical angular momentum on the sphere, which starts to rotate. The situation here is reminiscent of one of the versions of the so called ``Feynman paradox" (see page 175 in Feynman's book \cite{Feynman-Vol-II}), which is the statement that static electromagnetic fields may carry angular momentum. A related incarnation was discussed in Refs. \cite{Sharma,Griffiths} (see also Example 15.4 in the textbook \cite{zangwill2013modern}) and considers a ferromagnetic charged conducting sphere. The electromagnetic field of such a system carries an angular momentum. Above the Curie temperature the sphere demagnetizes and the initial angular momentum is converted to a mechanical one, and the sphere starts to rotate \cite{Griffiths}. 

In the case considered here, the magnetization is not spontaneous, but induced by the electric field originating from the point charge via the axion term in Eq.~(\ref{Eq:Axion-ED}). Changes in the magnetization occur by varying the electric field,
\begin{equation}
  \label{Eq:H}
  \vec{H}=-4\pi\frac{\partial\mathcal{L}}{\partial\vec{B}}=\frac{1}{\mu}\vec{B}+\frac{\alpha\theta}{\pi}\vec{E}=\vec{B}-4\pi\vec{M},
\end{equation} 
achieved by changing the position of the point charge outside the spherical TI. In view of the above equation, we therefore expect that also in the TI case the variation of the magnetization would cause the TI body to rotate. More specifically, since there are no external current source, the induced current at the surface due to the axion term is derived directly from Eq.~(\ref{Eq:H}) as
\begin{equation}
	\label{Eq:j-axion}
	\vec{j}=-\frac{\alpha c}{4\pi^2}\nablab\theta\times\vec{E}, 
\end{equation} 
which has the typical form of a Hall current. 
For a spherical TI of radius $a$ in vacuum, $\theta(R)=\theta H(a-R)$, where $H(x)$ is the Heaviside function, so the surface current becomes, 
\begin{eqnarray}
	\label{Eq:j-Hall}
	\vec{j}=\frac{\alpha \theta c} {4 \pi^2}\delta(R-a) \hat{\vec{R}} \times \vec{E}.
\end{eqnarray} 
Thus, an external point charge $q$ placed at a distance $d>a$ induces an electric field profile that in turn causes the Hall current (\ref{Eq:j-Hall}) on the surface of this sphere. Consequently, a magnetic field is generated. 

In the present context, the reader might collate this work and the well-known Einstein-de Haas (EdH) effect, but we should emphasize the difference. In Ref. \cite{Sharma}, the sphere rotates regardless of the intrinsic structure of the constituents. Moreover, there are two ways to induce the rotation: either by demagnetization or by discharging the body. This is similar to the present work's setup.

In this paper we solve the problem of a point charge in the presence of a spherical TI, which here is viewed as a dielectric sphere with axion electrodynamics and an induced surface Hall current. As usual, the main role played by the axion term in this case is affecting the boundary conditions of the problem \cite{Qi_image_monopole,Nogueira-van_den_Brink_PhysRevResearch.4.013074}. Since $\theta(R)=\theta={\rm const}$ inside the dielectric sphere while vanishing outside, the Maxwell equations do not modify, since the axion term in the Lagrangian (\ref{Eq:Axion-ED}) can actually be shown to be a total derivative \cite{Qi-2008}. The unusual result following from these considerations is that electric fields induce magnetic fields and vice-versa even in an inherently static situation. For this reason, there must be an angular momentum being generated as well within the static framework, in some sense is a more extreme realization of the scenario underlying the Feynman paradox. After solving the static Maxwell equations in detail Sec. II, we proceed with the calculation of the angular momentum for this spherical TI in Sec. III, where we also argue that varying the distance $d$ of the charge $q$ transfers the angular momentum of the electromagnetic field to the sphere in the form of a mechanical one, causing it to rotate. In Sec. III we also derive an exact expression for the induced angular momentum due to the Hall current of electrons moving on the surface. This allows to derive the expression of the electronic velocity as a function of $\theta$ and $a/d$.

\section{Point charge in the presence of a dielectric sphere} 

\begin{figure}
	\begin{center}
		\includegraphics[width=8cm]{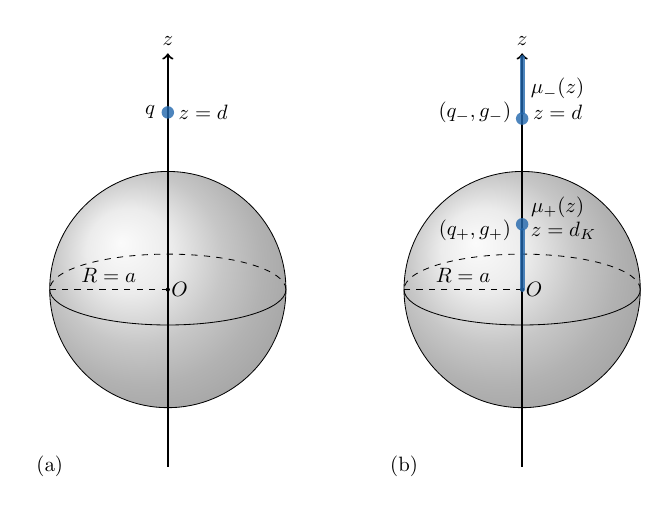}
	\end{center}
	\caption{(a) Schematic representation of the problem of a spherical TI of radius $R=a$ in the presence of a point charge $q$ located 
		at $\vec{R}_0=(0,0,d)$. (b) Due to the axion term, the point charge induces both electric and magnetic fields which can be thought 
		to be generated by pairs of electric and magnetic point charges $(q_-,g_-)$ and $(q_+,g_+)$ located at $z=d$ (the so called Kelvin point) and 
		$z=d_K=a^2/d$, respectively, and magnetic line charge densities $\mu_\pm(z)$ (shown in blue) in the intervals $0<z<d_K$ and 
		$z>d$.}
	\label{Fig:sph-TI}
\end{figure}

In the problem for a point charge in the presence of a dielectric sphere 
the image electric charges are not given solely by point charges. 
Besides the usual reflected image charge at the so called Kelvin point $d_K=a^2/d$ ($a$ is the radius of the sphere and $d>a$ is the position of the point charge $q$ along the $z$-axis), there is also a line charge density extending from the center of the 
sphere to $d_K$ \cite{Lindell,Norris,Norris-Comment}.  Likewise, the usual transmitted image charge given by a point charge at $\vec{R}_0=(0,0,d)$ has to be supplemented by a line charge density extending from $\vec{R}_0$ to infinity along the $z$-axis. 
This solution has been first obtained by Neumann \cite{neumann1883hydrodynamische} in the late nineteenth century and rediscovered by Lindell \cite{Lindell,Lindell-AJP,Bussemer}.  The solution for a TI sphere is obtained by accounting for the modified boundary conditions implied by the axion term. This leads  us to consider magnetic image charges as well, since the axion term induces a magnetic field in view of the magnetoelectric coupling it entails. The situation is illustrated in Fig.~\ref{Fig:sph-TI}. 

Since in the static problem the fields $\vec{E}$ and $\vec{H}=\vec{B}+(\alpha\theta/\pi)\vec{E}$ have vanishing curls, it is better to consider  the 
static Maxwell equations in the form
\begin{equation}
\nablab\cdot\left\{\left[\epsilon+\left(\frac{\alpha\theta}{\pi}\right)^2\right]\vec{E}-\frac{\alpha\theta}{\pi}\vec{H}\right\}=
4\pi\rho,
\end{equation}
\begin{equation}
\nablab\cdot\left(\vec{H}-\frac{\alpha\theta}{\pi}\vec{E}\right)=0,
\end{equation}
where $\rho(\vec{R})=q\delta^3(\vec{R}-\vec{R}_0)$.  
Since $\nablab\times\vec{E}=0$ and $\nablab\times\vec{H}=0$, we can introduce 
the scalar potentials $\phi$ and $\chi$ such that $\vec{E}=-\nablab\phi$ and $\vec{H}=-\nablab\chi$. The symmetry of the problem 
implies that all potentials depend only on $R=|\vec{R}|$ and $\Theta$ in spherical coordinates $(R,\Theta,\varphi)$ (we are 
using $\Theta$ instead of $\theta$ to avoid confusion with the axion field). Defining $\phi_\pm$ and $\chi_\pm$ as 
the potentials outside (+) and inside (-) the sphere, the boundary conditions are simply given by
\begin{equation}
\phi_+(R=a,\Theta)=\phi_-(R=a,\Theta), 
\end{equation}
\begin{equation}
\chi_+(R=a,\Theta)=\chi_-(R=a,\Theta), 
\end{equation}
\begin{equation}
\left[\epsilon+\left(\frac{\alpha\theta}{\pi}\right)^2\right]\left.\frac{\partial\phi_-}{\partial R}\right|_{R=a}- 
\left.\frac{\partial\phi_+}{\partial R}\right|_{R=a}=\frac{\alpha\theta}{\pi}\left.\frac{\partial\chi_-}{\partial R}\right|_{R=a},
\end{equation}
\begin{equation}
\left.\frac{\partial\chi_-}{\partial R}\right|_{R=a}-\left.\frac{\partial\chi_+}{\partial R}\right|_{R=a}
=\frac{\alpha\theta}{\pi}\left.\frac{\partial\phi_-}{\partial R}\right|_{R=a}.
\end{equation}
In the Supplemental Material of Ref. \cite{Qi_image_monopole} it was assumed that $\vec{B}$, rather than $\vec{H}$, is given by 
the gradient of a potential, which led additionally to a boundary condition stating the discontinuity of the tangential 
components of $\vec{B}$ due to the axion term. This does not happen in our case because we have $\nablab\times\vec{H}=0$, leading 
naturally to the choice $\vec{H}=-\nablab\chi$. As a consequence, 
\begin{equation}
\vec{B}=\vec{H}-\frac{\alpha\theta}{\pi}\vec{E}=-\nablab\chi+\frac{\alpha\theta}{\pi}\nablab\phi,
\end{equation}
implying obviously  that $\vec{B}=\vec{H}=-\nablab\chi$ for $R>a$, but not inside the sphere. Thus, if we introduce a magnetic 
scalar potential $\lambda$ such that $\vec{B}=-\nablab\lambda$, we have 
\begin{equation}
\lambda_+=\chi_+,
\end{equation}
\begin{equation}
\lambda_-=\chi_--\frac{\alpha\theta}{\pi}\phi_-,
\end{equation}
outside and inside the spherical TI, respectively. 
The electric and magnetic potentials are given by the standard expansions in terms of Legendre polynomials 
\cite{stratton2007electromagnetic}. The electric potential outside the sphere has the standard form
\begin{equation}
\label{Eq:phi+}
\phi_+(R,\Theta)=\frac{q}{|\vec{R}-d\hat{\bf z}|}+\sum_{n=0}^\infty \frac{b_n}{R^{n+1}}P_n(\cos\Theta). 
\end{equation}
For $a<R< d$ we expand the contribution from the point charge alone to obtain, 
\begin{equation}
\label{Eq:phi+expansion-R<d}
\phi_+(R,\Theta)=\sum_{n=0}^{\infty}\left[\frac{q}{d}\left(\frac{R}{d}\right)^n+
\frac{b_n}{R^{n+1}}\right]P_n(\cos\Theta),
\end{equation}
while for $R>d$ we have
\begin{equation}
\label{Eq:phi+expansion-R>d}
\phi_+(R,\Theta)=\frac{1}{R}\sum_{n=0}^{\infty}(qd^n+b_n)\frac{P_n(\cos\Theta)}{R^n}.
\end{equation}
Furthermore, the following expansions hold
\begin{equation}
\phi_-(R,\Theta)=\sum_{n=0}^{\infty}a_nR^nP_n(\cos\Theta),
\end{equation}
\begin{equation}
\chi_+(R,\Theta)=\sum_{n=0}^\infty\frac{\beta_n}{R^{n+1}}P_n(\cos\Theta),
\end{equation}
\begin{equation}
\chi_-(R,\Theta)=\sum_{n=0}^\infty\alpha_nR^nP_n(\cos\Theta).
\end{equation}
Using the boundary conditions we easily obtain
\begin{equation}
\label{Eq:an}
a_n=\frac{(2n+1)^2q/d^{n+1}}{(2n+1)[n(1+\epsilon)+1]+n(n+1)(\alpha\theta/\pi)^2},
\end{equation}

\begin{equation}
\label{Eq:bn}
b_n=\frac{nqa^{2n+1}}{d^{n+1}}\frac{(2n+1)(1-\epsilon)-(n+1)(\alpha\theta/\pi)^2}{(2n+1)[n(1+\epsilon)+1]+n(n+1)(\alpha\theta/\pi)^2},
\end{equation}
\begin{equation}
\label{Eq:alphan}
\alpha_n=\frac{n(2n+1)(\alpha\theta/\pi)}{(2n+1)[n(1+\epsilon)+1]+n(n+1)(\alpha\theta/\pi)^2}\frac{q}{d^{n+1}},
\end{equation}
\begin{equation}
\label{Eq:betan}
\beta_n=a^{2n+1}\alpha_n.
\end{equation}

As in the problem with a plane geometry \cite{Nogueira-van_den_Brink_PhysRevResearch.4.013074}, the smallness of $(\alpha\theta/\pi)^2$ causes  
electric potentials to not differ appreciably from 
the ones for a dielectric sphere in the presence of a point charge for $\theta=0$. The magnetic scalar potentials, on the other hand, 
are proportional to $\alpha\theta/\pi$ in leading order. 

Given the above coefficients of the Legendre polynomial expansion of the potentials, we obtain the magnetic scalar potentials, 
\begin{widetext}
	\begin{equation}
	\label{Eq:la+}
	\lambda_+(R,\Theta)=\frac{q\alpha\theta}{\pi}\sum_{n=0}^\infty
	\frac{n(2n+1)}{(2n+1)[n(1+\epsilon)+1]+n(n+1)(\alpha\theta/\pi)^2}\frac{a^{2n+1}}{d^{n+1}}
	\frac{P_n(\cos\Theta)}{R^{n+1}},
	\end{equation}
	\begin{equation}
	\label{Eq:la-}
	\lambda_-(R,\Theta)=-\frac{q\alpha\theta}{\pi}\sum_{n=0}^\infty
	\frac{(n+1)(2n+1)}{(2n+1)[n(1+\epsilon)+1]+n(n+1)(\alpha\theta/\pi)^2}\frac{R^{n}}{d^{n+1}}
	P_n(\cos\Theta). 
	\end{equation}
\end{widetext}
The above result for $\lambda_\pm$ agrees with the one obtained in the Supplemental Saterial of Ref. \cite{Qi_image_monopole} in 
the special case where the permeabilities $\mu_1=\mu_2=1$, $\epsilon_1=1$, and $\epsilon_2=\epsilon$. Later, during the discussion 
of the image charges Ref. \cite{Qi_image_monopole} assumes in addition that $\epsilon=1$. For the problem with $\theta=0$ this 
was called the regime of a "vanishing sphere" by Lindell \cite{Lindell}, since this leads to a vanishing of the reflection image electric charge. 
This is readily seen from Eq.~(\ref{Eq:bn}), which vanishes when $\epsilon=1$ and $\theta=0$.  However, the transmission image  
electric charge does not vanish in this regime, a result at odds with the one in Ref. \cite{Qi_image_monopole}. Thus, although the 
magnetic potentials above agree with the result of Ref. \cite{Qi_image_monopole}, the electric potentials disagree. 
This error ended affecting the result of the main text in a planar geometry, which 
was obtained as the limit case of a very large radius $a$ of the sphere, keeping the distance 
$|d-a|\ll a$ between the charge and the surface of the sphere constant. This discrepancy has also been noted by Karch \cite{Karch}. Indeed, 
it is easy to see that the result in Ref. \cite{Qi_image_monopole} for the case of a planar geometry leads to a vanishing of both 
transmission and reflection image electric charges when $\epsilon_1=\epsilon_2=1$ and $\theta=0$. 

We also note that the expressions for the scalar potentials in Sec. IVD of Ref. \cite{Martin-Ruiz_PhysRevD.94.085019} are fully consistent with our analysis. However, the authors use the method of Green's functions and treat the region $a<r<d$ implicitly, providing the calculation for only two regions, $a<d<R$ and $R<a<d$. Applying an appropriate Legendre polynomial expansion to the Coulomb potential allows to show that it agrees with our work.

The reflection and transmission 
image magnetic line density charges $\mu_i^\pm(z)$ can be determined by comparing the expansions (\ref{Eq:la+}) and (\ref{Eq:la-}) with 
\begin{equation}
\lambda_+(R,\Theta)=\sum_{n=0}^\infty\int_{0}^{a}dz'\mu_i^+(z')\frac{(z')^n}{R^{n+1}}P_n(\cos\Theta),
\end{equation}
\begin{equation}
\lambda_-(R,\Theta)=\sum_{n=0}^\infty\int_{a}^{\infty}dz'\mu_i^-(z')\frac{R^n}{(z')^{n+1}}P_n(\cos\Theta),
\end{equation}
respectively. Therefore, $\mu_i^\pm(z)$ are determined by the integral equations, 
\begin{widetext}
	\begin{equation}
	\label{Eq:mu+}
	\int_{0}^{a}dz'\mu_i^+(z')\left(\frac{z'}{d_K}\right)^n=\frac{a}{d}\frac{q\alpha\theta}{\pi}\frac{n(2n+1)}{(2n+1)[n(1+\epsilon)+1]+n(n+1)(\alpha\theta/\pi)^2},
	\end{equation}
	\begin{equation}
	\label{Eq:mu-}
	\int_{a}^{\infty}dz'\mu_i^-(z')\left(\frac{d}{z'}\right)^{n+1}=-\frac{q\alpha\theta}{\pi}\frac{(n+1)(2n+1)}{(2n+1)[n(1+\epsilon)+1]+n(n+1)(\alpha\theta/\pi)^2},
	\end{equation}
	where $d_K=a^2/d$ is the Kelvin image point. 
	
	It is possible to determine analytically the image magnetic charge line densities 
	to leading order in $\alpha\theta/\pi$, since in this case we have simply, 
	\begin{equation}
	\int_{0}^{a}dz'\mu_i^+(z')\left(\frac{z'}{d_K}\right)^n=\frac{a}{d}\frac{q\alpha\theta}{\pi}\frac{n}{n(1+\epsilon)+1}
	+{\cal O}\left[\left(\frac{\alpha\theta}{\pi}\right)^3\right],
	\end{equation}
	\begin{equation}
	\int_{a}^{\infty}dz'\mu_i^-(z')\left(\frac{d}{z'}\right)^{n+1}=-\frac{q\alpha\theta}{\pi}\frac{n+1}{n(1+\epsilon)+1}
	+{\cal O}\left[\left(\frac{\alpha\theta}{\pi}\right)^3\right]. 
	\end{equation}
\end{widetext}
For $\theta=\pi$ we have that $(\alpha\theta/\pi)^3\approx 3.9\times 10^{-7}$, so the analytical result is  
essentially exact. 

For the reflection image we can use the identity \cite{Lindell-AJP}
\begin{equation}
\int_{0}^adz\frac{d}{dz}\left[\left(\frac{z}{d_K}\right)^{n_0}H(d_K-z)\right]\left(\frac{z}{d_K}\right)^n=-\frac{n}{n+n_0},
\end{equation}
where $H(z)$ denotes the Heaviside step function. Thus, to leading order, 
\begin{equation}
\mu_i^+(z)=-\frac{a}{d}\frac{q\alpha\theta}{\pi(1+\epsilon)}\frac{d}{dz}\left[\left(\frac{z}{d_K}\right)^{\frac{1}{1+\epsilon}}
H(d_K-z)\right],
\end{equation}
and therefore,
\begin{eqnarray}
\mu_i^+(z)&=&\frac{a}{d}\frac{q\alpha\theta}{\pi(1+\epsilon)}\delta(z-d_K)
\nonumber\\
&-&\frac{a}{dd_K}\frac{q\alpha\theta}{\pi(1+\epsilon)^2}\left(\frac{z}{d_K}\right)^{-\frac{\epsilon}{1+\epsilon}}H(d_K-z).
\end{eqnarray}
Thus, the reflection image magnetic density charge consists of a point image magnetic charge $aq\alpha\theta/[\pi d(1+\epsilon)]$ at 
the Kelvin point $z=d_K$ plus an image magnetic charge line density along the interval $0<z<d_K$. Note that,   
\begin{equation}
	\label{Eq:Neutral}
	\int_{0}^a dz\mu_i^+(z)=0. 
\end{equation}

For the transmission image magnetic charge density we have to use the identity \cite{Lindell-AJP},
\begin{eqnarray}
&&\int_{a}^\infty dz\left[A\delta(z-d)+B\left(\frac{d}{z}\right)^{n_0}H(z-d)\right]\left(\frac{d}{z}\right)^{n+1}
\nonumber\\
&=&A+\frac{Bd}{n+n_0}. 
\end{eqnarray}
This yields 
\begin{eqnarray}
\mu_i^-&=&-\frac{q\alpha\theta}{\pi(1+\epsilon)}\delta(z-d)
\nonumber\\
&-&\frac{\epsilon}{(1+\epsilon)^2}\frac{q\alpha\theta}{\pi d}\left(\frac{d}{z}\right)^{\frac{1}{1+\epsilon}}H(z-d),
\end{eqnarray}
corresponding to an image magnetic point charge at $z=d$ plus an image magnetic line charge density extending over the line $z>d$. 

A similar analysis can be easily made for the electric potential. However, at leading order in $\alpha\theta/\pi$ it does not differ 
appreciably from the well known result for an ordinary dielectric sphere  \cite{neumann1883hydrodynamische,Lindell,Lindell-AJP}. \\

\section{Angular momentum} 

\subsection{The angular momentum from Hall currents}
\label{App:L-mech}



Let us calculate the angular momentum solely due to the mechanical motion of charge carriers on the surface of the spherical TI, which is encoded in the Hall current of Eq.~(\ref{Eq:j-Hall}). 
%
Since
\begin{eqnarray}
	\hat{\vec{R}} \times \vec{E}=E_ {\Theta} \hat{\vec{\varphi}}
\end{eqnarray}
the current density is along the $\varphi$ direction,
\begin{eqnarray}
	j_{\varphi}=\frac{\alpha \theta c}{4 \pi^2} \delta(R-a) E_{\Theta}(a,\Theta)
\end{eqnarray}
From it we infer the momentum density,
\begin{eqnarray}
	p_\varphi=\frac{m_e j_{\varphi}}{e},
\end{eqnarray}
where $m_e$ is the mass of the electrons of charge $e$ moving on the surface. Hence, the angular momentum density is given by
\begin{eqnarray}
	\vec{l}_{\text {Mech}}&=&\frac{m_e}{e}({\vec{R}} \times \vec{j})
	\nonumber\\
	&=&-\frac{m_e c}{e} \frac{\alpha \theta}{4 \pi^2} \delta(R-a) a E_{\Theta}(a, \Theta) \hat{\vec{\Theta}}.
\end{eqnarray}
Only the $z$ component of the above expression will contribute to the angular momentum when integrated over the whole space, 
\begin{eqnarray}
	l_{\text{Mech},z}=\frac{m_e c \alpha \theta a}{4 \pi^2 e} \delta(R-a) E_\Theta(a, \Theta) \sin\Theta.
\end{eqnarray}
Next, we need, 

\begin{equation}
	E_{\Theta}(a,\Theta)=\frac{1}{a} \sum_{n=0}^{\infty}\left[\frac{q}{d}\left(\frac{a}{d}\right)^n+\frac{b_n}{a^{n+1}}\right] \frac{\partial}{\partial \Theta} P_n(\cos\Theta),
\end{equation}
and therefore, 

\begin{widetext}
	\begin{eqnarray}
		\label{Eq:Lmech}
		L_{\text {Mech}, z}&=&\int d^3 R l_{\text {Mech},z}=\frac{\alpha \theta c a^2 m_e}{2 \pi e} \sum_{n=0}^{\infty}\left[\frac{q}{d}\left(\frac{a}{d}\right)^n+\frac{b_n}{a^{n+1}}\right] \int_0^\pi d \Theta \sin ^2 \Theta \frac{\partial}{\partial \Theta} P_n(\cos \Theta)\\ \nonumber
		&=&-\frac{2 \alpha \theta c a^2}{3 \pi e}\left(\frac{qa}{d^2}+\frac{b_1}{a^2}\right)
		=-\frac{6 N\alpha \theta c m_e}{\pi} \frac{a^3}{ d^2} \frac{1}{3(2+\epsilon)+2(\alpha\theta/\pi)^2}=I_e\omega_e(\epsilon,d,\theta),
	\end{eqnarray}
\end{widetext} 
where $I_e=Nm_e a^2$ is the moment of inertia, and we have used
\begin{eqnarray}
	b_1= \frac{qa^3}{d^2} \frac{3(1-\epsilon)-2(\alpha \theta / \pi)^2}{3(2+\epsilon)+2(\alpha\theta/\pi)^2}
\end{eqnarray}
Equation~(\ref{Eq:Lmech}) defines the angular velocity
\begin{equation}
	\omega_e(\epsilon,d,\theta)=-\frac{6\alpha\theta a c}{\pi d^2}\frac{1}{3(2+\epsilon)+2(\alpha\theta/\pi)^2}.
\end{equation}

We can express Eq.~(\ref{Eq:Lmech}) in the form
\begin{equation}
	\label{Eq:LzMech_redef}
	L_{\text{Mech}, z}=-N m_e v_{e} a
\end{equation}
where $m_e$ is the mass of the electron, and  $v_{e}$ is the velocity of the surface electrons, which is given explicitly by
\begin{equation}
	v_e=\frac{\alpha\theta}{\pi}\frac{6c}{3(2+\epsilon)+2(\alpha\theta/\pi)^2}\left(\frac{a}{d}\right)^2.
\end{equation} 
Electron states on the TI surface are topologically protected ensuring no back scattering \cite{RevModPhys.82.3045}. Thus, it is useful to compare the numerical value of the above velocity to its typical values on the surface of actual materials. Inserting, for instance, $\theta=\pi$, $a=50$ nm, $d=51$ nm, and $\epsilon=20$, we obtain, $v_e \approx 1.91 \times 10^7$ $\text{cm}\cdot\text{s}^{-1}$. This value is quite reasonable and well in the range of expected values for the electron velocities on TI surfaces (for example, Bi$_2$Se$_3$ has $v_e\approx 5\times 10^{7}~\text{cm}\cdot\text{s}^{-1}$ \cite{Bi2Se3-velocity}). 
The 1 nm separation between the point charge and the surface of the sphere is consistent with the continuum description of axion electrodynamics, within which the boundary conditions arise naturally and are applied infinitesimally close to the surface.
\subsection{Angular momentum of the electromagnetic field}

We now turn on to the calculation of the angular momentum of the electromagnetic field for such a system. In contrast to (\ref{Eq:Lmech}), the latter cannot be calculated in closed form analytically, since the result will be an infinite series that cannot be cast into an easy functional form. 
 
Since electromagnetic momentum density is given by 
the well known textbook formula, $\Pib=(\vec{E}\times\vec{B})/(4\pi c)$,  the angular momentum carried by the electromagnetic field is obtained by integrating over 
all space the angular momentum density, $({\bf r}+z\hat{\bf z})\times\Pib$. 
The components $L_x$ and $L_y$ vanish due rotational invariance. Hence,  
we have,  
\begin{widetext}
    \begin{eqnarray}
\label{Eq:Lz-sphere}
L_z&=&\frac{1}{2c}\int_a^\infty dR\int_{0}^{\pi}d\Theta R^2\sin^2\Theta\left(\frac{\partial\phi_+}{\partial R}\frac{\partial\lambda_+}{\partial\Theta}
-\frac{\partial\phi_+}{\partial \Theta}\frac{\partial\lambda_+}{\partial R}\right)
\nonumber\\
&+&\frac{1}{2c}\int_0^a dR\int_{0}^{\pi}d\Theta R^2\sin^2\Theta\left(\frac{\partial\phi_-}{\partial R}\frac{\partial\lambda_-}{\partial\Theta}
-\frac{\partial\phi_-}{\partial \Theta}\frac{\partial\lambda_-}{\partial R}\right).
\nonumber\\
\end{eqnarray}

The integral corresponding to the potentials outside the sphere has to be split into two contributions, one for the region $a<R<d$ and another one for the region $R>a$, due to the expansions (\ref{Eq:phi+expansion-R<d}) and (\ref{Eq:phi+expansion-R>d}).  
Similarly to the planar case \cite{Nogueira-van_den_Brink_PhysRevResearch.4.013074}, it is easy to show that the second integral in Eq.~(\ref{Eq:Lz-sphere}) corresponding to the interior of the TI vanishes and thus we actually have
\begin{align}
\label{Eq:Lz-sphere-1}
L_z=\frac{1}{2c}\int_a^\infty dR\int_{0}^{\pi}d\Theta R^2\sin^2\Theta\left(\frac{\partial\phi_+}{\partial R}\frac{\partial\lambda_+}{\partial\Theta}
-\frac{\partial\phi_+}{\partial \Theta}\frac{\partial\lambda_+}{\partial R}\right).
\end{align}
More explicitly, 
\begin{eqnarray}
L_z&=&-\frac{q}{2cd}\sum_{n,m=0}^\infty\frac{\beta_m}{d^n}\int_{a}^{d}dR\int_{0}^\pi d\Theta\sin^3\Theta R^{n-m}
\left[nP_n(\cos\Theta)P_m'(\cos\Theta)+(m+1)P_m(\cos\Theta)P_n'(\cos\Theta)\right]
\nonumber\\
&+&\frac{q}{2c}\sum_{n,m=0}^\infty (n+1)\left(\beta_md^n-\beta_nd^m\right)\int_d^\infty \frac{dR}{R^{m+n+1}}\int_{0}^\pi d\Theta\sin^3\Theta 
P_n(\cos\Theta)P_m'(\cos\Theta)
\nonumber\\
&+&\frac{1}{2c}\sum_{n,m=0}^\infty(n+1) \left(b_n\beta_m-b_m\beta_n\right)\int_a^\infty  \frac{dR}{R^{m+n+1}}\int_{0}^\pi d\Theta\sin^3\Theta 
P_n(\cos\Theta)P_m'(\cos\Theta),
\nonumber
\end{eqnarray}
where the prime denotes a derivative with respect to $\cos\Theta$. The above expression can be simplified further by using the orthogonality 
of Legendre polynomials along with the 
identities 
\begin{equation}
\sin^2\Theta P_m'(\cos\Theta)=m\left[P_{m-1}(\cos\Theta)-\cos\Theta P_m(\cos\Theta)\right],
\end{equation}
\begin{equation}
\cos\Theta P_n(\cos\Theta)=\frac{(n+1)P_{n+1}(\cos\Theta)+nP_{n-1}(\cos\Theta)}{2n+1}.
\end{equation}
Therefore, we get
\begin{eqnarray}
\label{eq:IntegerLz}
L_z=(N\alpha)^2\hbar\Upsilon(\epsilon,d/a),
\end{eqnarray}
where $N=q/e \in \mathbb{N}$ and
\begin{eqnarray}
\label{eq:LzSum}
\Upsilon(\epsilon,d/a)=\sum_{n=0}^{\infty}\bigg[\frac{\mathcal{B}_{n-1}}{d^{n-1}}\left(1-\frac{a^2}{d^2}\right)\frac{n^2}{1-4n^2}
+\bigg(1+\frac{B_n}{d^n}\bigg)
\bigg(\frac{\mathcal{B}_{n+1}}{d^{n+1}}\frac{(n+1)(n+2)}{(2n+1)(2n+3)}+\frac{\mathcal{B}_{n-1}}{d^{n-1}}\frac{n(n+1)}{1-4n^2}\bigg)\bigg],
\end{eqnarray}
\end{widetext}
with $\mathcal{B}_n=\beta_n/q\alpha$ and $B_n=b_n/q$. We will use the latter constants to leading order in $\alpha$, since there is already a prefactor $\alpha^2$ in Eq.~(\ref{eq:IntegerLz}). We have, 
\begin{equation}
	\mathcal{B}_n=\frac{a^{2n+1}}{d^{n+1}}\frac{n\theta}{\pi[n(1+\epsilon)+1]}+\mathcal{O}(\alpha^2),
\end{equation}
\begin{equation}
	B_n=\frac{a^{2n+1}}{d^{n+1}}\frac{n(1-\epsilon)}{n(1+\epsilon)+1}+\mathcal{O}(\alpha^2).
\end{equation}

\begin{figure}[ht]
	\includegraphics[width=8cm]{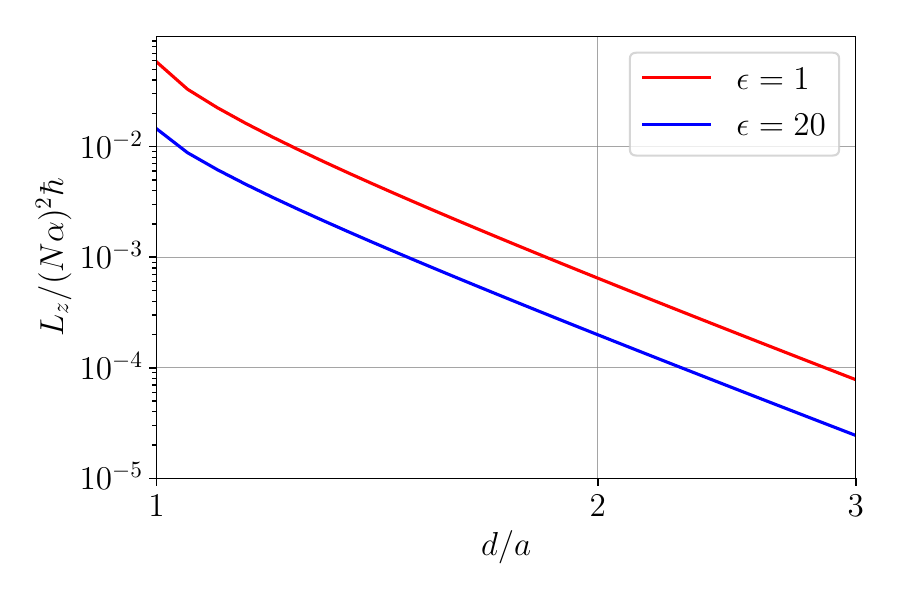}
	\caption{Result of the summation in Eq.~(\ref{eq:LzSum}) for the $L_z/(N\alpha)^2\hbar$ dependence on $d/a$ in log-log scale for  $\epsilon=1 \text{ and } 20$. $N$ is taken to be 1.}
	\label{Fig:loglogLz}
\end{figure}

Note that the angular momentum (via the function $\Upsilon$) is only a function of $d/a$, although this is not immediately obvious from the series of Eq.~(\ref{eq:LzSum}). 
The angular momentum as a function of $d/a$ and $\theta=\pi$ is shown in Fig. \ref{Fig:loglogLz}. 
\textcolor{black}{It is interesting to observe that $L_z$ can be sizable via e.g., $N$. Unlike Eq.~(\ref{Eq:Lmech}), the dependence on $N$ is not linear, but quadratic.}
Thus, in a typical STM tip one usually has $N\sim 10^2$ \cite{STM}, implying that actually measuring the angular momentum should not represent a major technical challenge.
For $d\gg a$, when the charge is very far away from the TI, Eq.~(\ref{eq:LzSum}) turns into
\begin{equation}
	L_z^\infty\approx \frac{4(N\alpha)^2\hbar}{3(2+\epsilon)}\left(\frac{a}{d}\right)^3.
\label{Eq:Lz_short}
\end{equation}
\begin{figure}[ht]
	\includegraphics[width=8cm]{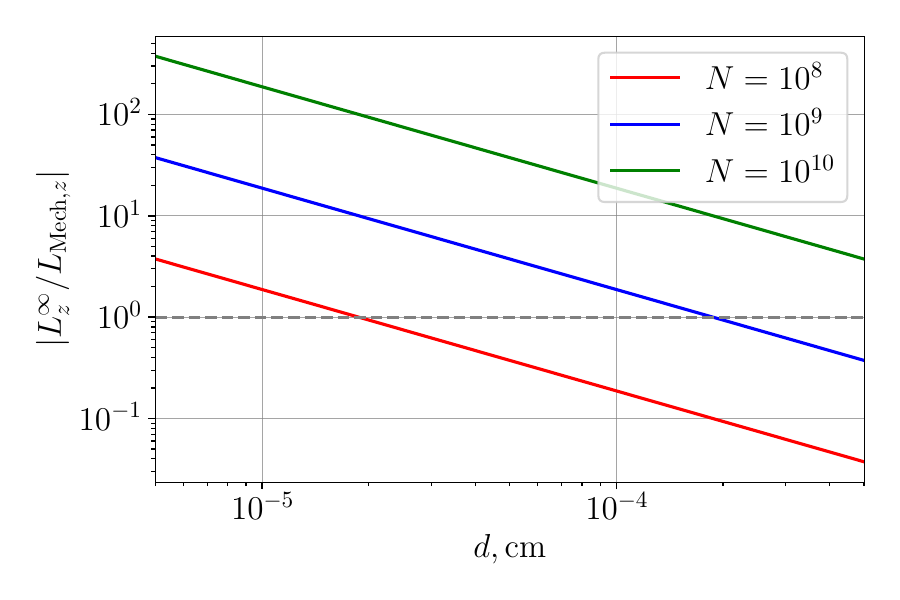}
	\caption{The $d$ dependence of the ratio of the electromagnetic angular momentum $L_z^\infty$ in Eq.~(\ref{Eq:Lz_short}) when $d\gg a$, to the angular momentum of the Hall current $L_{\text{Mech},z}$, with different values of $N$. The gray dashed line crosses points, where the ratio is equal to unity.}
	\label{Fig:lratio}
\end{figure}

Varying the position of the point charge immediately causes a variation in $L_z$, which in view of the conservation of angular momentum should be compensated by a mechanical rotation of the spherical TI. 

 We can compare the numerical values of electromagnetic $L_z$ and $L_{\text{Mech}, z}$ due to the azimuthal Hall current, which are opposite in sign. Rewriting $L_{\text{Mech}, z}=-N^2 \alpha^2 \hbar \Delta$, with $\Delta$ being a dimensionless quantity comparable to $\Upsilon$, and using $N=10^2$ as a typical STM value, we calculate $\Upsilon\approx 1.14 \times 10^{-2}$ and $\Delta \approx 1.56 \times 10^{-1}$, which yields one order of magnitude difference.

In Fig.~\ref{Fig:lratio} we show the $d$ dependence of the ratio of the electromagnetic angular momentum in Eq.~(\ref{Eq:Lz_short}) to the one in Eq.~(\ref{Eq:Lmech}). Though the analytical expressions in the above mentioned equations are not identical, we can determine a regime where the ratio is equal to unity numerically. In first order of $\alpha$ this ratio depends only on $N$ and $d$, and does not on either $\epsilon$ or $a$. Varying the former ones, we get the family of points corresponding to the transfer of angular momentum to the surface Hall current.

Finally, it is worth to mention that Eq.~(\ref{Eq:Lz_short}) includes only odd powers of the fine-structure constant, while Eq.~(\ref{Eq:Lmech}) only even powers of $\alpha$.

We may wonder what happens in the so called `ìnfinite sphere" limit, where the radius $a$ of the sphere is much larger than $d$, as discussed in Ref. \cite{Lindell-AJP}. This case would correspond to the one of a semi-infinite TI, whose angular momentum due to the electromagnetic field was calculated in Ref. \cite{Nogueira-van_den_Brink_PhysRevResearch.4.013074}. The result turns out to be independent of the distance of the charge $q$ to the surface, in a situation reminiscent of the angular momentum of a dyon (a dipole consisting of an electric charge and a magnetic monopole). Indeed, the angular momentum of a dyon is independent of its size. However, for a slab of finite thickness, the dependence on the distance of the charge to the surface arises in the angular momentum. An example of such a calculation is found in the Appendix.

\section{Conclusion}
Using the effective axion field theory approach and method of image charges, in this work, we solve a problem for static fields induced by a point charge near a spherical TI. As a result, the static field profiles have been found as well as the expression for the electromagnetic angular momentum and its power law dependence on the distance $d$ of the point charge from the center of the spherical TI body. As elucidated along the way, changing $d$ converts the  angular momentum stored in the electromagnetic field into a mechanical one, as required by its fundamental law of conservation. In addition, the external point charge induces azimuthal Hall currents on the surface, which also carry angular momentum, so varying $d$ also changes its value. From the latter we derive an expression for the induced velocity of an electron on the surface of the spherical TI as a function of $\theta$ and $a/d$.  

We also point out the difference between responses to the presence of the point charge of a trivial insulator and the TI in the same geometry. In this paper, the response features the reflection and transmission magnetic line charge densities $\mu^\pm(z)$, in addition to the electric line charge densities extending throughout the same regions.

\begin{acknowledgments}
	We acknowledge financial support by the Deutsche Forschungsgemeinschaft (DFG, German Research Foundation), through SFB 1143 Project A5 and the W{\"u}rzburg-Dresden Cluster of Excellence on Complexity and Topology in Quantum Matter-ct.qmat (EXC 2147, Project ID No. 390858490), as well as support by Bundesministerium für Forschung, Technologie und Raumfahrt (BMFTR) funding through Project No. 01DK24008 (GU-QuMat). 
\end{acknowledgments}

\appendix
\section{Infinite slab of thickness \texorpdfstring{$L$}{TEXT}}
\label{appendix:A}

\begin{figure}[h]
	\includegraphics[width=8cm]{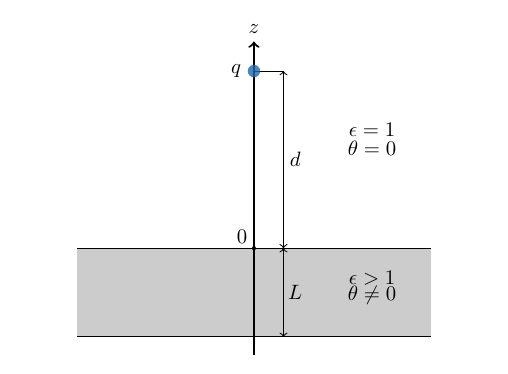}
	\caption{Infinite magnetoelectric slab of thickness $L$ discussed in Appendix.}
	\label{Fig:infslab}
\end{figure}

For semi-infinite magnetoelectric discussed in detail in Ref. \cite{Nogueira-van_den_Brink_PhysRevResearch.4.013074}, the $z$ component of electromagnetic angular momentum turns out to not depend on the distance $d$ from the magnetoelectric surface to the point-charge placed above,
\begin{align}
\label{eq:monopoleL}
L_z=-N^2 \frac{\Phi}{\Phi_0} \frac{\hbar}{2},
\end{align}
where the flux
\begin{align}
\Phi=\frac{\alpha^2 \theta \Phi_0}{\pi(1+\epsilon+\kappa)},
\end{align}
and $\Phi_0=hc/e$ is the flux quantum, $\kappa=(1/2)(\alpha\theta/\pi)^2$.
Now, to investigate the origin of the $d$ dependence of the angular momentum, we turn to an intermediate example of an infinite slab of thickness $L$ (Fig.~\ref{Fig:infslab}), since such geometry can be viewed as stereographic projection of the sphere onto the plane. In this case, the north and south poles become the upper and lower parts of the slab, respectively.

The system of equations in part IV of Ref. \cite{Nogueira-van_den_Brink_PhysRevResearch.4.013074} is extended by one region $z<-L$,
\begin{eqnarray}
-\frac{d^2 \hat{\phi}}{d z^2}+p^2 \hat{\phi}(\boldsymbol{p},z)&=&4 \pi \delta(z-d), \quad z>0\\ \nonumber
-\epsilon \frac{d^2 \hat{\phi}}{d z^2}+\epsilon p^2 \hat{\phi}(\boldsymbol{p}, z)&=&0,\quad -L<z<0\\
-\frac{d^2 \hat{\phi}}{d z^2}+p^2 \hat{\phi}(\boldsymbol{p}, z)&=&0, \quad z<-L\nonumber,
\end{eqnarray}
and we get two additional boundary conditions:
\begin{align}
\begin{aligned}\setcounter{mysubequations}{0}
    \text{\mysubnumber}\quad &\hat{\phi}(\boldsymbol{p}, z=-\eta)=\hat{\phi}(\boldsymbol{p}, z=+\eta)\\
    \text{\mysubnumber}\quad &\left.\frac{d \hat{\phi}}{d z}\right|_{z=+\eta}-\left.\epsilon \frac{d \hat{\phi}}{d z}\right|_{z=-\eta}=x|\boldsymbol{p}|\hat{\phi}\left( \boldsymbol{p}, z=0\right)\\
  	\text{\mysubnumber}\quad &\hat{\phi}(\boldsymbol{p},z=d-\eta)=\hat{\phi}(\boldsymbol{p}, z=d+\eta)\\
	\text{\mysubnumber}\quad &\left.\frac{d \hat{\phi}}{d z}\right|_{z=d-\eta}-\left.\frac{d \hat{\phi}}{d z}\right|_{z=d+\eta}=4 \pi q\\
	\text{\mysubnumber}\quad &\hat{\phi}(\boldsymbol{p}, z=-L-\eta)=\hat{\phi}(\boldsymbol{p}, z=-L+\eta)\\
	\text{\mysubnumber}\quad &\left.\frac{d \hat{\phi}}{d z}\right|_{z=-L-\eta}-\left.\epsilon \frac{d \hat{\phi}}{d z}\right|_{z=-L+\eta}=-\kappa|\boldsymbol{p}|\hat{\phi}(\boldsymbol{p},z=-L).\\
    \end{aligned}
    \label{eq:boudarycond}
\end{align}
Hence, the ansatz
\begin{align}
\hat{\phi}(\boldsymbol{p},z) =
    \begin{cases}
      & D e^{-|\boldsymbol{p}| z},\quad z>d\\
      & B e^{|\boldsymbol{p}| z}+C e^{-|\boldsymbol{p}| z},\quad 0<z<d\\
      & F e^{|\boldsymbol{p}| z}+G e^{-|\boldsymbol{p}| z},\quad -L<z<0 \\
		& A e^{|\boldsymbol{p}| z}, \quad z<-L
    \end{cases}
\label{eq:slabansatz}
\end{align}
Using Eq.~(\ref{eq:boudarycond}) and Eq.~(\ref{eq:slabansatz}), we solve for the coefficients,
\begin{widetext}
\begin{align}
  \begin{split}
    B&=\frac{2 \pi q}{|\boldsymbol{p}|} e^{-|\boldsymbol{p}| d},\\
    A&=-\frac{8 \pi q}{|\boldsymbol{p}|} \frac{\epsilon e^{|\boldsymbol{p}|(2 L-d)}}{(1+\kappa-\epsilon)^2-(1+x\kappa+\epsilon)^2 e^{2 |\boldsymbol{p}|L}},\\
    F&=-\frac{4 \pi q}{|\boldsymbol{p}|} \frac{(1+\kappa+\epsilon) e^{|\boldsymbol{p}|(2 L-d)}}{(1+\kappa-\epsilon)^2-(1+\kappa+\epsilon)^2 e^{2 |\boldsymbol{p}| L}},
  \end{split}
\quad\quad
  \begin{split}
    G&=\frac{4 \pi q}{|\boldsymbol{p}|} \frac{(1+\kappa-\epsilon) e^{-|\boldsymbol{p}| d}}{(1+\kappa-\epsilon)^2-(1+\kappa+\epsilon)^2 e^{2 |\boldsymbol{p}| L}},\\
    C&=\frac{4 \pi q}{|\boldsymbol{p}|} e^{-|\boldsymbol{p}| d}\left[\frac{1+\kappa-\epsilon-(1+\kappa+\epsilon) e^{2 |\boldsymbol{p}| L}}{(1+\kappa-\epsilon)^2-(1+\kappa+\epsilon)^2 e^{2 |\boldsymbol{p}| L}}-\frac{1}{2}\right],\\
	D&=\frac{4\pi q}{|\boldsymbol{p}|}\left\{\sinh (|\boldsymbol{p}| d)+e^{-|\boldsymbol{p}| d}\left[\frac{1+\kappa-\epsilon-(1+\kappa+\epsilon) e^{2 |\boldsymbol{p}| L}}{(1+\kappa-\epsilon)^2-(1+\kappa+\epsilon)^2 e^{2 |\boldsymbol{p}| L}}\right]\right\}.
  \end{split}
\end{align}
\end{widetext}

From the form of the above coefficients we see that the angular momentum will be $d$ dependent as a result of the finite thickness $L$. 

\bibliography{screening}

\end{document}